\documentclass[12pt]{article}
\usepackage{latexsym}
\usepackage{amsmath}
\usepackage{amsfonts}
\usepackage{amsthm}
\usepackage{amssymb}
\usepackage{epsfig}
\usepackage{graphicx}

\catcode`ð=\active
 \defð{\u{g}}
 \catcode`Ð=\active
 \defÐ{\u{G}}
 \catcode`Ý=\active
 \defÝ{\. I}
 \catcode`ö=\active
 \defö{\"{o}}
 \catcode`Ö=\active
 \defÖ{\"O}
 \catcode`ü=\active
 \defü{\"{u}}
 \catcode`Ü=\active
 \defÜ{\"{U}}
 \catcode`Þ=\active
 \defÞ{\c{S}}
 \catcode`þ=\active
 \defþ{\c{s}}
 \catcode`ý=\active
 \defý{{\i}}
 \catcode`ç=\active
 \defç{\d{c}}
 \catcode`Ç=\active
 \defÇ{\d{C}}
\def\rar{\rightarrow}

\begin{document}
\begin{center}
\bf The Forward-Backward Asymmetry in the\\ $B \rightarrow
\pi\ell^+\ell^-$ Decay
\end{center}

\begin{center}
Altuð ARDA\footnote{arda@hacettepe.edu.tr} \\ Hacettepe
University, Physics Education, 06532, Ankara, Turkey
\end{center}

\vspace*{1cm}

\begin{abstract}
Using the most general effective Hamiltonian comprising
scalar,vector and tensor type interactions, we have written the
branching ratio, the forward-backward (FB) asymmetry and the
normalized FB asymmetry as functions of the new Wilson
coefficients. It is found that the branching ratio depends on all
new coefficients,but the dependence of asymmetries on coefficients
could be analyzed only for one Wilson coefficient.
\end{abstract}

\newpage

\section{Introduction}
The flavour-changing neutral current(FCNC) processes provide an
excellent testing ground for the Standard Model(SM), and are
possibly the most sensitive to the various extensions to the SM,
because these transitions occur at the loop level in the SM. Among
all the FCNC phenomena, the rare decays of B-mesons are especially
important~\cite{Chetyrkin}, since one can both test the SM and
search for the new physics beyond it.\\
The rare B-decays induced by $b \rar s(d)\ell^+\ell^-$ transitions
at quark level have received a lot of attention, since they can be
used to determine the parameters of the SM, such as the
Cabibbo-Kobayashi-Maskawa (CKM) matrix elements~\cite{Aliev1}.
Many useful observables, such as branching ratio, CP-asymmetry,
lepton asymmetry etc., in the decays of B-mesons induced by $b
\rar s(d)\ell^+\ell^-$ transitions has been investigated in the
literature in the framework of the SM and its
extensions~\cite{Goto,Fukae,Ali1,Aliev2,Aliev3,Kruger,Erkol,Iltan}.
Another interesting observable is the forward-backward asymmetry,
because the forward-backward asymmetry may become zero for the
certain value of the dilepton invariant mass for the $B \rar
K^*\ell^+\ell^-$ decay~\cite{Burdman}. On the other hand the
lepton asymmetry is zero for the exclusive $B \rar K\ell^+\ell^-$
decay within the SM~\cite{Burdman,Ali2}. This quantity and its
zero position have been also investigated for different decays
within the SM and its various
extensions~\cite{Arda,Ghinculov,Gao,Chen}. It has been found that
the new Wilson coefficients could contribute to the zero position
of the forward-backward asymmetry and the zero position is most
sensitive to the coefficients~\cite{Arda}.\\ In the present work,
following to~\cite{Aliev1},we study the forward-backward asymmetry
in the exclusive $B \rar \pi\ell^+\ell^-$ decay,using the most
general effective Hamiltonian which has new Wilson coefficients
and analyze the
contributions coming from these coefficients within this model.\\
The organization of this work as follows. In Section II, starting
from the most general effective Hamiltonian, we compute the
differential decay width of the $B \rar \pi\ell^+\ell^-$ decay,and
the forward-backward asymmetry for this decay. In Section III, we
carry out the numerical analysis to study the dependence of the
asymmetry on the new Wilson coefficients and finally we conclude
in Section IV.

\section{Decay Width and Forward-Backward\\ Asymmetry}
The semileptonic decay $b \rar d\ell^+\ell^-$ is described,by
effective Hamiltonian,at the quark level as:
\begin{eqnarray}
{\cal
H}_{eff}=\,\frac{4G_F}{\sqrt{2}}\,V_{tb}V_{td}^*\,\sum_{i=1}^{10}C_i(\mu){\cal
O}_i(\mu),
\end{eqnarray}
where the full set of the operators ${\cal O}_i(\mu)$ in the SM
are given in~\cite{Grinstein}. The effective coefficient of the
operator ${\cal O}_9(\mu)$ can be defined as:
\begin{eqnarray}
C_9^{eff}(\hat{s})&=&\,C_9+g(z,
\hat{s})\,(3C_1+C_2+3C_3+C_4+3C_5+C_6)\nonumber\\&-&\,\frac{1}{2}\,g(1,
\hat{s})\,(4C_3+4C_4+3C_5+C_6)-\,\frac{1}{2}\,g(0,
\hat{s})\,(C_3+3C_4)\nonumber\\&+&\,\frac{1}{2}\,(3C_3+C_4+3C_5+C_6).
\end{eqnarray}
Here, $\hat{s}=q^2/m^2_B$ where q is the momentum transfer and
$z=m_c/m_b$. The functions $g(z, \hat{s}), g(1, \hat{s})$ and
$g(0, \hat{s})$ can be found in~\cite{Kim}.\\
Neglecting the d quark mass, the above Hamiltonian leads to the
following matrix element for the $b \rar d\ell^+\ell^-$ decay:
\begin{eqnarray}
{\cal
M}&=&\,\frac{G_F\alpha}{\sqrt{2}\pi}\,V_{tb}V^*_{td}\Bigg[(C^{eff}_9-C_{10})\bar{d}_L\gamma_{\mu}b_L
\bar{\ell}_L\gamma^{\mu}\ell_L\nonumber\\&+&(C^{eff}_9+C_{10})\bar{d}_L\gamma_{\mu}b_L
\bar{\ell}_R\gamma^{\mu}\ell_R\nonumber\\&-&2C^{eff}_7\hat{m}_b\bar{d}i
\sigma_{\mu\nu}\frac{\hat{q}^{\nu}}{\hat{s}}Rb\,\hat{\ell}\gamma^{\mu}\ell\Bigg],
\end{eqnarray}
Where, $L/R=(1\mp \gamma^5)/2$, $\hat{m}_b=m_b/m_B$ and
$b(d)_{L,R}=[(1\mp \gamma^5)/2]b(d)$.\\
The matrix element for the $b \rar d\ell^+\ell^-$ decay coming
from the most general effective Hamiltonian reads as~\cite{Aliev4}
\begin{eqnarray}
{\cal
M}&=&\,\frac{G_F\alpha}{\sqrt{2}\pi}\,V_{tb}V^*_{td}\Bigg[C_{LL}\bar{d}_L\gamma_{\mu}b_L
\bar{\ell}_L\gamma^{\mu}\ell_L+C_{LR}\bar{d}_L\gamma_{\mu}b_L
\bar{\ell}_R\gamma^{\mu}\ell_R\nonumber\\&+&C_{RL}\bar{d}_R\gamma_{\mu}b_R
\bar{\ell}_L\gamma^{\mu}\ell_L+C_{RR}\bar{d}_R\gamma_{\mu}b_R
\bar{\ell}_R\gamma^{\mu}\ell_R\nonumber\\&+&C_{LRLR}\bar{d}_Lb_R
\bar{\ell}_L\ell_R+C_{RLLR}\bar{d}_Rb_L
\bar{\ell}_L\ell_R\nonumber\\&+&C_{LRRL}\bar{d}_Lb_R
\bar{\ell}_R\ell_L+C_{RLRL}\bar{d}_Rb_L
\bar{\ell}_R\ell_L\nonumber\\&+&C_T\bar{d}\sigma_{\mu\nu}b\bar{\ell}
\sigma^{\mu\nu}\ell+iC_{TE}\bar{d}\sigma_{\mu\nu}b\bar{\ell}
\sigma_{\alpha\beta}\ell\epsilon^{\mu\nu\alpha\beta}\Bigg].
\end{eqnarray}
The matrix element of the effective Hamiltonian over $\pi$ and B
meson states in the $B \rar \pi\ell^+\ell^-$ decay are
parametrized in terms of the form factors,and to calculate the
amplitude of the $B \rar \pi\ell^+\ell^-$ decay,we need
\begin{eqnarray}
\left< \pi\left|\bar{d}\gamma_{\mu}b\right|
B\right>&=&(P_{\mu}-\frac{1-\hat{m}^2_\pi}{\hat{s}}\,
q_{\mu})f_{+}+(\frac{1-\hat{m}^2_\pi}{\hat{s}})\, q_{\mu}f_{0}\,,
\end{eqnarray}
with $f_+(0)=f_0(0)$,
\begin{eqnarray}
\left< \pi\left|\bar{d}\sigma_{\mu\nu}b\right|
B\right>=-i(P_{\mu}q_{\nu}-P_{\nu}q_{\mu})\,\frac{f_T}{m_B+m_{\pi}}\,,
\end{eqnarray}
\begin{eqnarray}
\left< \pi\left|\bar{d}i\sigma_{\mu\nu}q^{\nu}b\right|
B\right>=\Big[P_{\mu}q^2-(m^2_B-m^2_{\pi})q_{\mu}\Big]\,\frac{f_T}{m_B+m_{\pi}}\,,
\end{eqnarray}
\begin{eqnarray}
\left< \pi\left|\bar{d}b\right|
B\right>=\,\frac{m_B(1-\hat{m}^2_\pi)}{\hat{m}_b}\,f_0\,,
\end{eqnarray}
where $P=p_1+p_2$, $p_1$ and $p_2$ are the four momenta of the B
and $\pi$ mesons, respectively.\\
Using the matrix elements (5)-(8), we obtain the amplitude for $B
\rar \pi\ell^+\ell^-$ decay as:
\begin{eqnarray}
{\cal M}&=&\,\frac{G_F\alpha}{\sqrt{2}\pi}\,V_{tb}V^*_{td}
\Bigg[A^{'}P_{\mu}(\bar{\ell}\gamma^{\mu}\ell)+B^{'}q_{\mu}(\bar{\ell}\gamma^{\mu}\ell)
+C^{'}P_{\mu}(\bar{\ell}\gamma^{\mu}\gamma_5\ell)\nonumber\\&+&
D^{'}q_{\mu}(\bar{\ell}\gamma^{\mu}\gamma_5\ell)+A(\bar{\ell}\ell)
+B(\bar{\ell}\gamma_5\ell)+iC(P_{\mu}q_{\nu}-P_{\nu}q_{\mu})(\bar{\ell}\sigma^{\mu\nu}\ell)
\nonumber\\&+&D(P_{\mu}q_{\nu}-P_{\nu}q_{\mu})(\epsilon^{\mu\nu\alpha\beta}
\bar{\ell}\sigma_{\alpha\beta}\ell)\Bigg]\,,
\end{eqnarray}
where
\begin{eqnarray}
A^{'}&=&(2C^{eff}_9+C_{LL}+C_{LR}+C_{RL}+C_{RR})f_{+}-4\hat{m}_bC^{eff}_7
\frac{f_T}{1+\hat{m}_\pi}\,, \nonumber \\
B^{'}&=&(2C^{eff}_9+C_{LL}+C_{LR} +C_{RL}+C_{RR})f_{-}
+4\hat{m}_bC^{eff}_7\frac{1-\hat{m}_\pi}{\hat{s}}f_T\,, \nonumber \\
C^{'}&=&\Big[2C_{10}+C_{LR}+C_{RR}-(C_{LL}+C_{RL})\Big]f_{+}\,, \nonumber \\
D^{'}&=&\Big[2C_{10}+C_{LR}+C_{RR}-(C_{LL}+C_{RL})\Big]f_{-}\,, \nonumber \\
A&=&(C_{LRLR}+C_{LRRL}+C_{RLLR}+C_{RLRL})\frac{m_B(1-\hat{m}^2_\pi)}{\hat{m}_b}f_0\,, \nonumber \\
B&=&\Big[C_{LRLR}+C_{RLLR}-(C_{LRRL}+C_{RLRL})
\Bigg]\frac{m_B(1-\hat{m}^2_\pi)}{\hat{m}_b}f_0\,, \nonumber \\
C&=&-4C_T\frac{f_T}{m_B+m_\pi}\,, \nonumber \\
D&=&4C_{TE}\frac{f_T}{m_B+m_\pi}\,.\nonumber
\end{eqnarray}
with
\begin{eqnarray}
f_-=\frac{1-\hat{m}^2_\pi}{\hat{s}}(f_0-f_+)\,. \nonumber
\end{eqnarray}
We would like to note that in calculating the double-decay width
we take into account of the massless case. However, the
calculation for the massive case is given in the Appendix. Using
the matrix element in Eq.(9), the double differential decay width
can be calculated as:
\begin{eqnarray}
\frac{d^2\Gamma}{d\hat{s}dcos\theta}&=&\,\frac{G^2_F\alpha^2}{2^{13}\pi^5}\,
|V_{tb}V^*_{td}|^2m^3_B\lambda^{1/2}(1,\hat{m}^2_{\pi},\hat{s})\nonumber\\
&\times&\Bigg[\Big[-\Big(\left|A^{'}\right|^2+\left|C^{'}\right|^2\Big)\lambda+4\lambda
\hat{s}m^2_B\Big(\left|C\right|^2+4\left|D\right|^2\Big)\Big]cos^2\theta\nonumber\\
&-&\Big[16Re\Big(D^{'}D^{*}\Big)+Re(AC^{*})+8Re(BD^{*})\Big]\lambda^{1/2}\hat{s}cos\theta
\nonumber\\&+&\Big(\left|A^{'}\right|^2+\left|C^{'}\right|^2\Big)\lambda+
\Big(\left|A\right|^2+\left|B\right|^2\Big)\,\frac{\hat{s}}{m^2_B}\Bigg]\,.
\end{eqnarray}
where
$\lambda(1,\hat{m}^2_{\pi},\hat{s})=1+\hat{m}^4_{\pi}+\hat{s}^2-2\hat{m}^2_{\pi}\hat{s}
-2\hat{m}^2_{\pi}-2\hat{s}$ and $\theta$ is the angle between the
four momenta of $\pi$ meson and $\ell^-$ in the dilepton
CMS-frame. The kinematical variables are bounded as
\begin{eqnarray}
-1\leq cos\theta\leq1\,,\nonumber\\
4\hat{m}^2_{\ell}\leq\hat{s}\leq(1-\hat{m}_{\pi})^2\, .\nonumber
\end{eqnarray}
From Eq.(10) one can easily obtain the single differential decay
rate
\begin{eqnarray}
\frac{d\Gamma}{d\hat{s}}&=&\,\frac{G^2_F\alpha^2}{2^{11}3\pi^5}\,
|V_{tb}V^*_{td}|^2m^3_B\lambda^{1/2}(1,\hat{m}^2_{\pi},\hat{s})\nonumber\\
&\times&\Bigg[\Big(\left|A^{'}\right|^2+\left|C^{'}\right|^2\Big)\lambda+
\hat{s}\Big[\frac{3}{2m^2_B}\Big(\left|A\right|^2+\left|B\right|^2\Big)+2\lambda
m^2_B\Big(\left|C\right|^2+4\left|D\right|^2\Big)\Big]\Bigg]\,.
\end{eqnarray}
On the other hand, the normalized asymmetry is defined as
\begin{eqnarray}
\frac{d\hat{A}_{FB}}{d\hat{s}}=\,\frac{dA_{FB}/d\hat{s}}{d\Gamma/d\hat{s}}=
\frac{\int_0^{1}(d^2\Gamma/d\hat{s}dcos\theta)-\int_{-1}^{0}(d^2\Gamma/d\hat{s}dcos\theta)}
{\int_0^{1}(d^2\Gamma/d\hat{s}dcos\theta)+\int_{-1}^{0}(d^2\Gamma/d\hat{s}dcos\theta)}\,,
\end{eqnarray}
where $dA_{FB}/d\hat{s}$ is the FB asymmetry and can be expressed
with the help of Eq.(10) as
\begin{eqnarray}
\frac{dA_{FB}}{d\hat{s}}&=&\,\frac{G^2_F\alpha^2}{2^{14}\pi^5}\,
|V_{tb}V^*_{td}|^2m^3_B\lambda^{1/2}(1,\hat{m}^2_{\pi},\hat{s})\hat{s}\nonumber\\
&\times&\Big[-32Re\Big(D^{'}D^*\Big)-2Re(AC^*)-16Re(BD^*)\Big]\,.
\end{eqnarray}
From Eqs.(11) and (13), one can see that the single differential
decay rate depends on also the square of all Wilson coefficients
near the other terms, but it is not the case for the FB asymmetry.
One notes that only the real part of the products of the auxiliary
functions A, B, C, D and $D^{'}$ contribute to the FB asymmetry,
that is the FB asymmetry is zero for the $B \rar \pi\ell^+\ell^-$
decay in the SM.
\section{Numerical Analysis}
The numerical values of the input parameters we used in our
analysis are the following:
\begin{eqnarray}
m_B=5.28\,GeV\,, m_{\pi}=0.14\,GeV\,, m_b=4.8\,GeV\,, m_c=1.4\,GeV\,,\nonumber\\
1/\alpha=129\,, G_F=1.6639\times10^{-5}\,GeV^2\,,
|V_{tb}V^*_{td}|=0.011\,.\nonumber
\end{eqnarray}
The values of the Wilson coefficients within the SM are given in
Table I, on the other hand we assume that all new Wilson
coefficients are real in the numerical analysis. In order to
complete the analysis we need the parametrization of the form
factors. For the values of that, we have used the results
of~\cite{Ball}. According to these results the form factors can be
parameterized as
\begin{eqnarray}
F(\hat{s})=\frac{F(0)}{1-a_F\hat{s}+b_F\hat{s}^2}\,.
\end{eqnarray}
where the parameters $F(0), a_F$ and $b_F$ for each form factor
are given in Table 2. \\
\begin{table}[ht]
$$
\begin{array}{|c c|| c c|}
\hline C_1 &-0.248 & C_6 &-0.031 \\
\hline
C_2 &1.107  &C_7  &-0.313 \\
\hline C_3 &0.011  &C_9 &4.344
\\ \hline C_4 &-0.026  &C_{10}
&-4.669 \\ \hline C_5 &0.007 &  & \\
\hline
\end{array}
$$
\caption{Values of the SM Wilson coefficients.}
\end{table}
\begin{table}[ht]
$$
\begin{array}{|l l l l|}
\hline
    & \phantom{-}F(0)  &\phantom{-}a_F  & \phantom{-}b_F \\ \hline
f_{+}&\phantom{-}0.305  &\phantom{-}1.29  &\phantom{-}0.206 \\
\hline f_0 &\phantom{-}0.305   &\phantom{-}0.266
&\phantom{-}-0.752
\\ \hline f_T &\phantom{-}0.296   &\phantom{-}1.28
&\phantom{-}0.193    \\ \hline
\end{array}
$$
\caption{Form factors for the $B \rar \pi$ transition.}
\end{table}
In our numerical analysis to investigate the dependence of the FB
asymmetry,the normalized FB asymmetry and the branching ratio to
the Wilson coefficients we consider two cases where one of the new
coefficients has two values $\pm |C_{10}|$, the others are set to
zero. From Eq.(13), after some calculations, one can see that all
terms which contribute to the FB asymmetry have the form of
product of any two new coefficients except one. According to this
result, we plot the dependence of the FB asymmetry on $\hat{s}$
for the $B \rar \pi\ell^+\ell^-$ decay for $C_{TE}=\pm |C_{10}|$
in Fig.(1). \\At first glance, we see that there is a linear
dependence between the FB asymmetry and the Wilson coefficient
$C_{TE}$ for these cases. When $C_{TE}=-|C_{10}|$, $C_{TE}$ gives
a positive contribution to the FB asymmetry.

\begin{figure}[htbp]
\vspace{-11.5cm}
\hspace{-3cm}
\includegraphics[height=13in,width=8in]{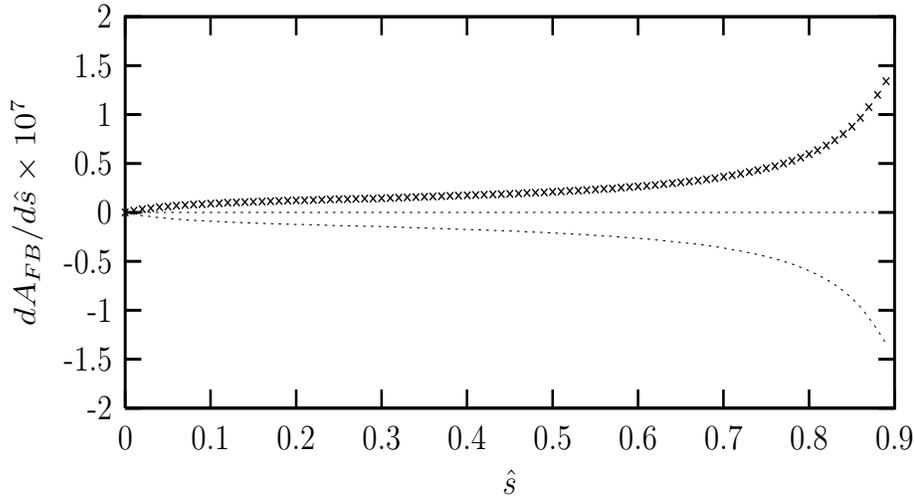}
\vspace*{-15cm}
\caption{The dependence of FB asymmetry on $\hat{s}$ for $B \rar
\pi\ell^+\ell^-$ decay corresponding to the cases
$C_{TE}=-|C_{10}|$(top curve) and $C_{TE}=+|C_{10}|$(bottom
curve).}
\end{figure}

\noindent However, when $C_{TE}=+|C_{10}|$ case, the contribution is negative. In both
cases,as $\hat{s}$ increases from 0.1 to 0.9, the contribution of
$C_{TE}$ increases in positive and negative direction. The
contributions of $C_{TE}=\pm |C_{10}|$ are symmetric with respect
to the zero axis.\\
With the help of Eq.(11) we plot the dependence of the branching
ratio of the $B \rar \pi\ell^+\ell^-$ decay to the new Wilson
coefficients with respect to $\hat{s}$. The results are shown in
Figs. (2)-(6). The contribution of $C_{RL}(C_{LR})$ to the
branching ratio is the same of $C_{LL}(C_{RR})$. The contributions
of $C_{LRRL}, C_{RLLR}$ and $C_{RLRL}$ are the same of $C_{LRLR}$.
Because of that we give only the figures for the new Wilson
coefficients $C_{LL}, C_{RR}, C_{LRRL}, C_{T}$ and $C_{TE}$. The
scalar and tensor type coefficients give the same contributions
for $+|C_{10}|$ and $-|C_{10}|$. These plots show that all new
coefficients give positive contributions to the branching ratio
and this observable is the most sensitive to $C_{LL}$.
In Fig.(7) we show the dependence of the normalized FB asymmetry
to the new Wilson coefficient $C_{TE}$. The normalized FB
asymmetry depends on one of the new coefficients like in the case
of the FB asymmetry. In this case, the dependence of the
normalized FB asymmetry is symmetric with respect to the zero axis
and $C_{TE}=-|C_{10}|$ gives a negative contribution,and
$C_{TE}=+|C_{10}|$ gives a positive one.

\begin{figure}[htbp]
\vspace{-12cm}
\hspace{-3cm}
\includegraphics[height=13in,width=8in]{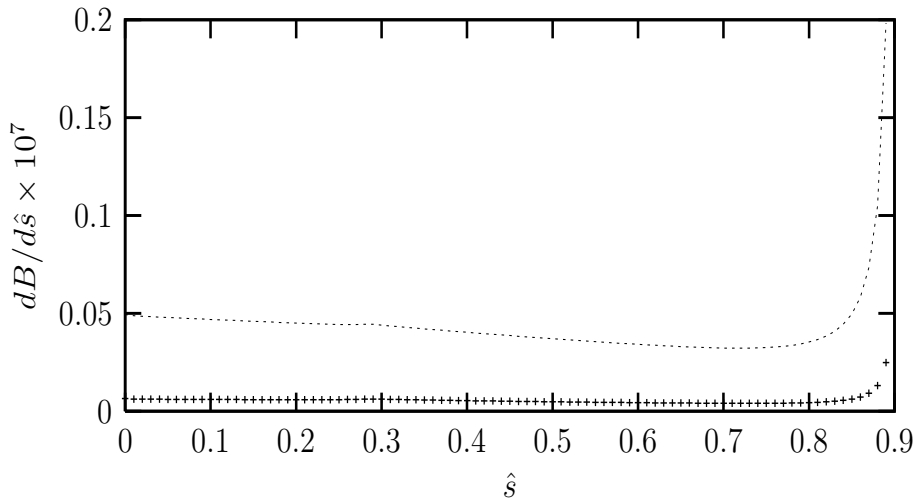}
\vspace{-15cm}
\caption{The dependence of the branching ratio on $\hat{s}$ for $B
\rar \pi\ell^+\ell^-$ decay corresponding to the cases
$C_{LL}=-|C_{10}|$(bottom curve) and $C_{LL}=+|C_{10}|$(top
curve).}
\end{figure}

\begin{figure}
\vspace{-10cm}
\hspace{-3cm}
\includegraphics[height=13in,width=8in]{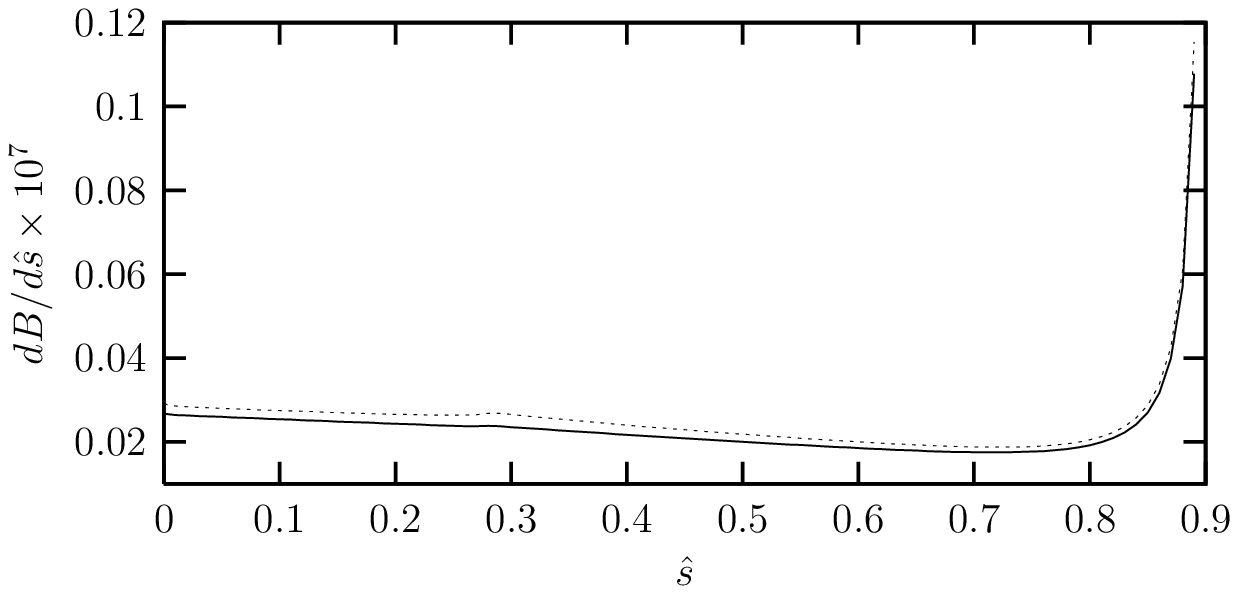}
\vspace{-15cm}
\caption{The dependence of the branching ratio on $\hat{s}$ for $B
\rar \pi\ell^+\ell^-$ decay corresponding to the cases
$C_{RR}=-|C_{10}|$(bottom curve) and $C_{RR}=+|C_{10}|$(top
curve).}
\end{figure}
\begin{figure}
\vspace{-10cm}
\hspace{-3cm}
\includegraphics[height=13in,width=8in]{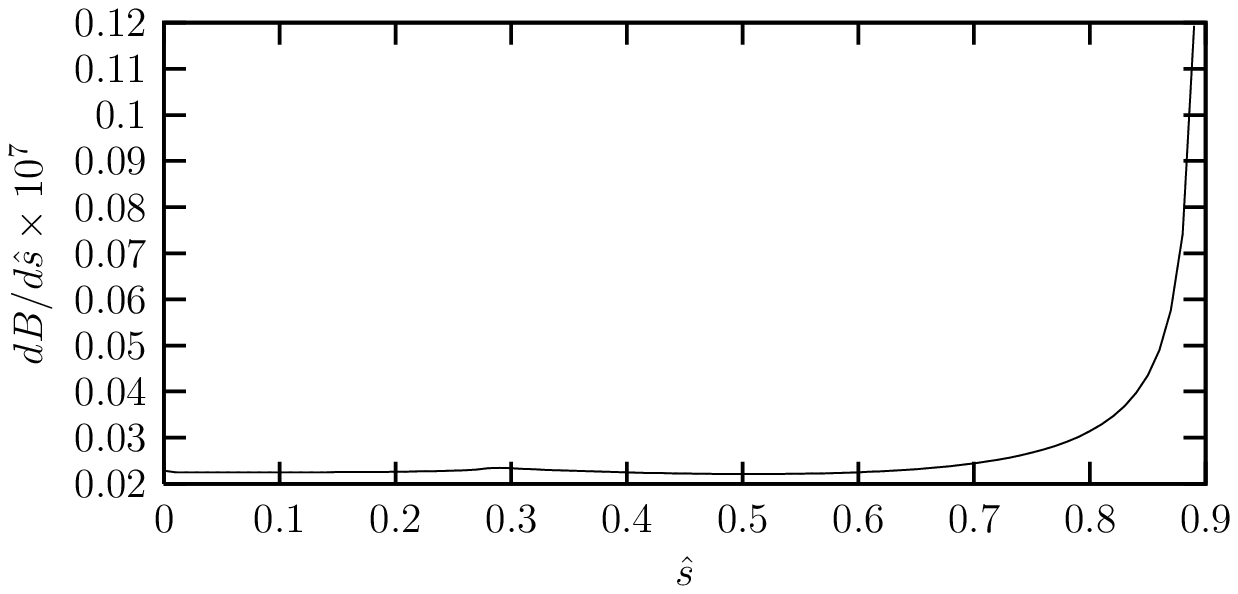}
\vspace{-15cm}
\caption{The dependence of the branching ratio on $\hat{s}$ for $B
\rar \pi\ell^+\ell^-$ decay corresponding to the cases
$C_{LRLR}=\pm |C_{10}|$.}
\end{figure}
\begin{figure}
\vspace{-10cm}
\hspace{-3cm}
\includegraphics[height=13in,width=8in]{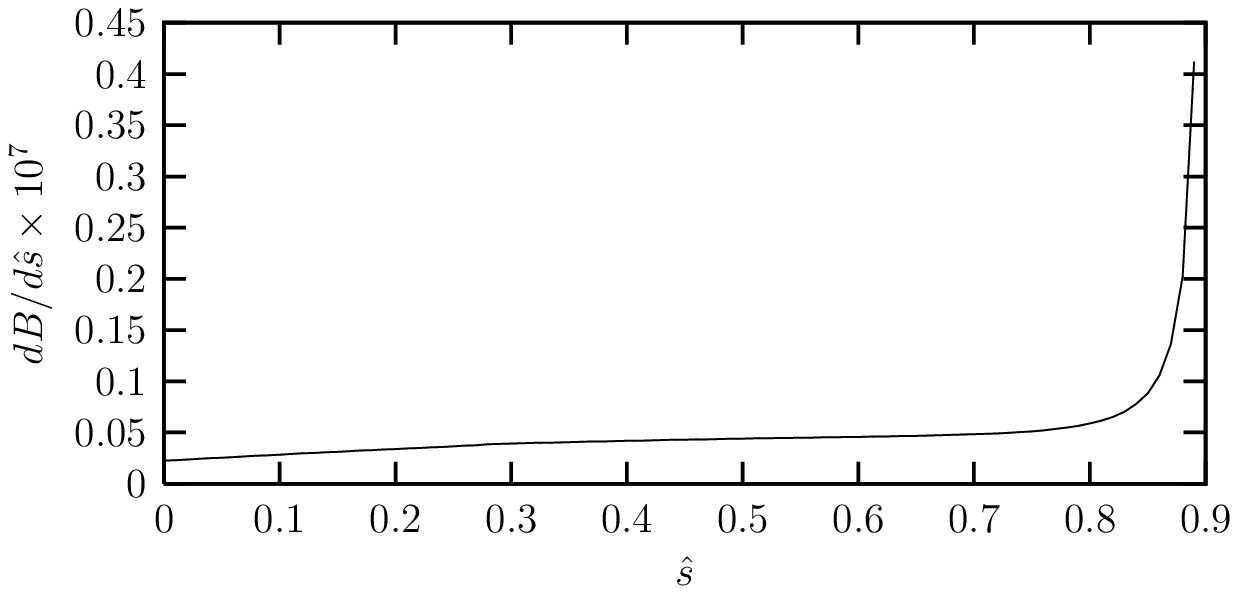}
\vspace{-15cm}
\caption{The dependence of the branching ratio on $\hat{s}$ for $B
\rar \pi\ell^+\ell^-$ decay corresponding to the cases $C_{T}=\pm
|C_{10}|$.}
\end{figure}
\begin{figure}
\vspace{-10cm}
\hspace{-3cm}
\includegraphics[height=13in,width=8in]{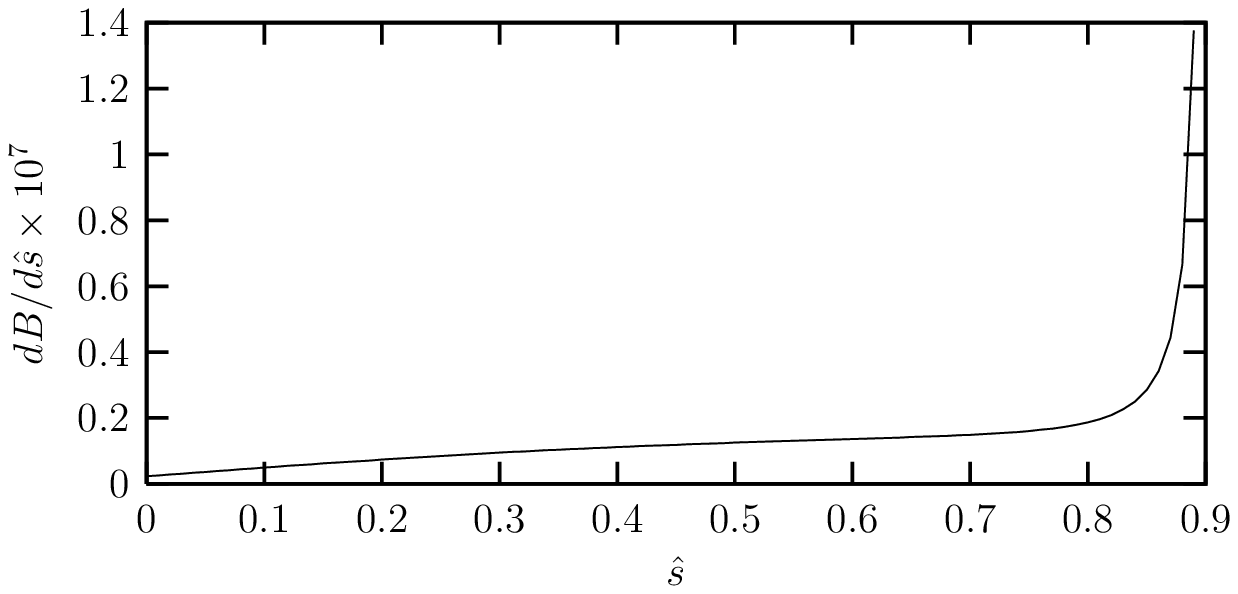}
\vspace{-15cm}
\caption{The dependence of the branching ratio on $\hat{s}$ for $B
\rar \pi\ell^+\ell^-$ decay corresponding to the cases $C_{TE}=\pm
|C_{10}|$.}
\end{figure}
\begin{figure}
\vspace{-10cm}
\hspace{-3cm}
\includegraphics[height=13in,width=8in]{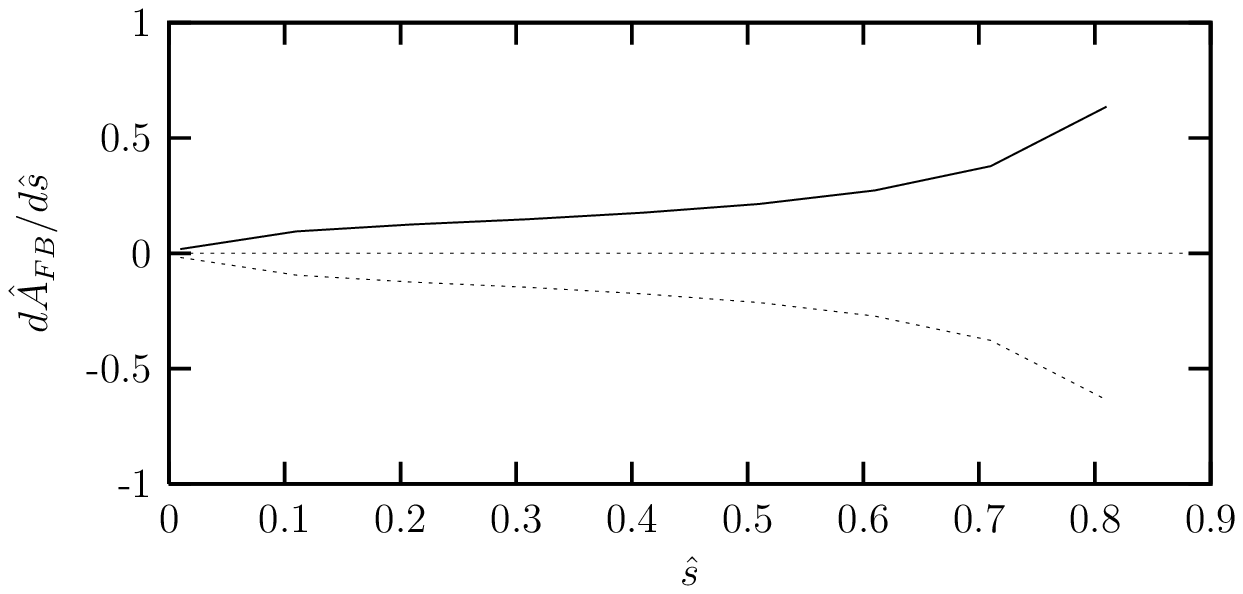}
\vspace{-15cm}
\caption{The dependence of the normalized FB asymmetry on
$\hat{s}$ for $B \rar \pi\ell^+\ell^-$ decay corresponding to the
cases $C_{TE}=-|C_{10}|$(top curve) and
$C_{TE}=+|C_{10}|$(bottom curve).}
\end{figure}
\section{Conclusion}
In the present work we have calculated the FB asymmetry, the
normalized FB asymmetry and the branching ratio within the most
general model of the $B \rar \pi\ell^+\ell^-$ decay. As we
expected, the FB asymmetry is zero for the $B \rar
\pi\ell^+\ell^-$ decay within the SM. We have analyzed the
dependence of these there observable to the new Wilson
coefficients coming from the most general model.\\ It is found
that there is a linearity between the FB asymmetry and the new
coefficient $C_{TE}$. The contribution coming from
$C_{TE}=-|C_{10}|$ is larger than that of $C_{TE}=+|C_{10}|$.\\
The normalized FB asymmetry depends on only the coefficient
$C_{TE}$. The contribution of $C_{TE}=-|C_{10}|$ is positive, on
the other hand, $C_{TE}=+|C_{10}|$ gives a negative contribution.
In both cases, the contributions are symmetric with respect to the
zero axis.\\
The branching ratio depends on all new Wilson coefficients,but
some of the contributions of the coefficients are the same with
each other. The branching ratio is the most sensitive to $C_{LL}$.
In vector type interactions, the coefficient $C_{TE}$ gives large
contribution  than that of $C_T$.
\section{Acknowledgments}
The author would like to thank Müge Boz for helpful discussions.\\

\newpage
\begin{center}
{\bf APPENDIX}
\end{center}
In the appendix we give the results for the massive lepton case.
First, we write the double differential decay width in terms of
the auxiliary functions with the same kinematical variables
\begin{eqnarray}
\frac{d^2\Gamma}{d\hat{s}dcos\theta}&=&\,\frac{G^2_F\alpha^2}{2^{11}\pi^5}\,
|V_{tb}V^*_{td}|^2m^3_B\lambda^{1/2}(1,\hat{m}^2_{\pi},\hat{s})\nonumber
\\&\times&\Bigg[\lambda\upsilon^2\Big[\hat{s}m^2_B
\Big(\left|C\right|^2+4\left|D\right|^2\Big)-\,\frac{1}{4}\,
\Big(\left|A^{'}\right|^2+\left|C^{'}\right|^2\Big)\Big]cos^2\theta\nonumber
\\&+&\lambda\Big[\frac{1}{4}\,
\Big(\left|A^{'}\right|^2+\left|C^{'}\right|^2\Big)+2m_{\ell}Re(A^{'}C^*)\Big]\nonumber
\\&+&\hat{m}^2_{\ell}\Big[\left|C^{'}\right|^2(2+2\hat{m}^2_{\pi}-\hat{s})+
\left|D^{'}\right|^2\hat{s}\Big]\nonumber
\\&+&(1-\hat{m}^2_{\pi})\Big[2\hat{m}^2_{\ell}Re(C^{'}D{'}^*)+\frac{\hat{m}_{\ell}}{m_B}
Re(C^{'}B^*)\Big]\nonumber
\\&+&\frac{\hat{s}}{m^2_B}\,\Big[\hat{m}_{\ell}Re(D^{'}B^*)+\frac{1}{4}\,
\Big(\left|B\right|^2+\upsilon^2\left|A\right|^2\Big)\Big]\nonumber
\\&+&\lambda\hat{s}m^2_B\left|C\right|^2(1-\upsilon^2)\nonumber \\
&-&\upsilon\lambda^{1/2}\Big[\frac{\hat{m}_{\ell}}{m_B}Re(A^{'}A^*)-4m_{\ell}
(\hat{m}^2_{\pi}-1)Re(C^{'}D^*)\nonumber
\\&+&\hat{s}\Big[4Re(D^{'}D^*)+2Re(DB^*)+\frac{1}{4}\,Re(AC^*)\Big]\Big]cos\theta\Bigg].
\nonumber
\end{eqnarray}
From this equation the single decay rate and the FB asymmetry is
given as follows
\begin{eqnarray}
\frac{d\Gamma}{d\hat{s}}&=&\,\frac{G^2_F\alpha^2}{2^{12}\pi^5}\,
|V_{tb}V^*_{td}|^2m^3_B\lambda^{1/2}(1,\hat{m}^2_{\pi},\hat{s})\nonumber
\\&\times&\Bigg[\lambda(1-\frac{1}{3}\,\upsilon^2)
\Big(\left|A^{'}\right|^2+\left|C^{'}\right|^2\Big)+
4\hat{m}^2_{\ell}\left|C^{'}\right|^2(2+2\hat{m}^2_{\pi}-\hat{s})\nonumber
\\&+&4\hat{m}^2_{\ell}\hat{s}\left|D^{'}\right|^2+\frac{\hat{s}}{m^2_B}\,
\Big(\left|B\right|^2+\upsilon^2\left|A\right|^2\Big)+8\hat{m}^2_{\ell}
(1-\hat{m}^2_{\pi})Re(C^{'}D{'}^*)\nonumber
\\&+&8m_{\ell}\lambda Re(A^{'}C^*)+4\frac{\hat{m}_{\ell}}{m_B}\,
(1-\hat{m}^2_{\pi})Re(C^{'}B^*)\nonumber
\\&+&\frac{4}{3}\,\lambda\hat{s}m^2_B\Big[3\left|C\right|^2+2\upsilon^2
\Big(2\left|D\right|^2-\left|C\right|^2\Big)\Big]\nonumber \\&+&
4\frac{\hat{m}_{\ell}}{m_B}\,\hat{s}Re(D^{'}B^*)\Bigg]\nonumber
\,,
\end{eqnarray}
\begin{eqnarray}
\frac{dA_{FB}}{d\hat{s}}&=&\,\frac{G^2_F\alpha^2}{2^{11}\pi^5}\,
|V_{tb}V^*_{td}|^2m^3_B\upsilon\lambda(1,\hat{m}^2_{\pi},\hat{s})\nonumber
\\&\times&\Bigg[4m_{\ell}(\hat{m}^2_{\pi}-1)Re(C^{'}D^*)-\frac{\hat{m}_{\ell}}{m_B}\,
Re(A^{'}A^*)\nonumber
\\&-&4\hat{s}Re(D^{'}D^*)-\frac{1}{4}\hat{s}Re(AC^*)-2\hat{s}Re(BD^*)\Bigg]\nonumber.
\end{eqnarray}
in the above equations
$\upsilon=\sqrt{1-\frac{4\hat{m}^2_{\ell}}{\hat{s}}}$\, is the
lepton velocity in the CMS-frame of leptons.

{000}

\end{document}